\documentstyle{elsart}
\begin{document}

\input FEYNMAN

ROM2F/95/10

\begin{frontmatter}

\title{\bf Deep Inelastic Scattering \\ \bf in Improved Lattice QCD \\
\it II. The second moment of structure functions\thanksref{cee}}
\thanks[cee]{Partially supported by EC Contract ``Computational
Particle Physics'', CHRX-CT92-0051}
\author[infn]{Giuseppe Beccarini},
\author[tovinfn]{Massimo Bianchi\thanksref{email3}},
\author[infn]{Stefano Capitani\thanksref{southampton}\thanksref{email1}}
and
\author[tovinfn]{Giancarlo Rossi\thanksref{email2}}
\address[infn]{INFN, Sezione di Roma 2, Via della Ricerca Scientifica,
I-00133 Roma, Italy}
\address[tovinfn]{Dipartimento di Fisica, Universit\`a degli Studi di
Roma ``Tor Vergata'', and INFN, Sezione di Roma 2,
Via della Ricerca Scientifica, I-00133 Roma, Italy}
\thanks[southampton]{Address from September 1995: Department of
Physics, University of Southampton, Highfield,
Southampton SO17 1BJ, United Kingdom}
\thanks[email1]{E-mail: capitani@vxrm70.roma1.infn.it}
\thanks[email2]{E-mail: rossig@vaxtov.roma2.infn.it}
\thanks[email3]{E-mail: bianchi@vaxtov.roma2.infn.it}

\begin{abstract}

In this paper we present the 1-loop perturbative computation of the
renormalization constants and mixing coefficients of the lattice
quark operators of rank three whose hadronic elements enter in the
determination of the second moment of Deep Inelastic Scattering (DIS)
structure functions.

We have employed in our calculations the nearest-neighbor improved
``clover-leaf'' lattice QCD action. The interest of using this action
in Monte Carlo simulations lies in the fact that all terms which in
the continuum limit are effectively of order $a$ ($a$ being the lattice
spacing) have been demonstrated to be absent from on-shell hadronic
lattice matrix elements. We have limited our computations to the
quenched case, in which quark operators do not mix with gluon operators.

We have studied the transformation properties under the hypercubic
group of the operators up to the rank five (which are related to moments
up to the fourth of DIS structure functions), and we discuss the choice
of the operators considered in this paper together with the feasibility
of lattice computations for operators of higher ranks.

To perform the huge amount of calculations required for the
evaluation of all the relevant Feynman diagrams, we have extensively
used the symbolic manipulation languages Schoonschip and Form.

\end{abstract}

\end{frontmatter}

\section{Introduction}

This is the second of two papers addressed to the problem of
computing in lattice QCD the 1-loop renormalization constants
and mixing coefficients of the operators of rank two and three
whose hadronic matrix elements respectively determine the first and
second moment of Deep Inelastic Scattering (DIS) structure functions.
In the first paper (hereafter referred to as \cite{first}) we have
reported the results for the case of the rank two operators.
Here we will extend these results to the quark operators of rank
three.

Let us start with a brief introduction to light-cone physics
and a summary of the improvement program in lattice gauge theories
\cite{Sym,Lus,She,rom1}. We refer the reader to \cite{first} or to
the review paper of Ref.~\cite{part} for more details.

\subsection{DIS and moments of the structure functions}

The quark operators whose matrix elements are related to the moments
of the DIS structure functions can be written in the form
\cite{gross,gross2,georgi}
\begin{displaymath}
O^{qS}_{\{\mu_{1} \cdots \mu_{N}\}} = \frac{1}{2^N}
\overline{\psi} \: \gamma_{\{ \mu_{1}} \!
\stackrel{\displaystyle \leftrightarrow}{D}_{\mu_{2}} \cdots
\stackrel{\displaystyle \leftrightarrow}{D}_{\mu_{N} \}}
(1 \pm \gamma_5) \psi
\end{displaymath}
\begin{equation}
O^{qNS}_{\{\mu_{1} \cdots \mu_{N}\}} = \frac{1}{2^N}
\overline{\psi} \: \gamma_{\{ \mu_{1}} \!
\stackrel{\displaystyle \leftrightarrow}{D}_{\mu_{2}} \cdots
\stackrel{\displaystyle \leftrightarrow}{D}_{\mu_{N} \}}
(1 \pm \gamma_5) \frac{\lambda^f}{2} \psi \label{eq:opsns} ,
\end{equation}
where the $\lambda^f$'s are flavor matrices. They are gauge invariant,
have twist two and are traceless and symmetrized with respect to all
Lorentz indices. $S$ and $NS$ superscripts refer to Singlet and
Non Singlet flavor structures.

In the unpolarized cross section the $\gamma_5$ contributions
present in Eqs.~(\ref{eq:opsns}) average to zero.
The other contributions have hadronic matrix elements of the form
\begin{equation}
\langle p| O^{(N)}_{\mu_{1} \cdots \mu_{N}} |p \rangle =
A_N(\mu) p_{\mu_{1}} \cdots p_{\mu_{N}}  + \mathrm{trace\ terms}
\label{eq:pppp},
\end{equation}
where $\mu$ is the subtraction point. They contain long distance
contributions (non-perturbative physics) and are related to
the $N$-th moment of the DIS structure functions by the equations
\cite{christ}
\begin{equation}
\left< x_B^N \right> = \int \d x_B \: x_B^N {\cal F}_{k}(q^{2},x_B)
= C_{N+1}(q^{2} / \mu^2) A_{N+1} (\mu)
\label{eq:momtwist}.
\end{equation}

\subsection{Improved lattice QCD}

The nearest-neighbor improved action (SW ``clover-leaf''
action) employed in these calculations is obtained by adding to the
standard Wilson action \cite{Wil} the Sheikholeslami--Wohlert \cite{She}
nearest-neighbor interaction term\footnote{Here $F_{n, \mu \nu}$ is not
the naive lattice ``plaquette''
\begin{equation}
P_{n, \mu \nu} = \frac{1}{2\mathrm{i}g_0a^2} (U_{n, \mu \nu}
- U^{+}_{n, \mu \nu}),~~~~~~U_{n, \mu \nu} =
U_{n,\mu} U_{n + \mu , \nu} U_{n + \nu , \mu}^{+} U_{n, \nu}^{+},
\label{eq:effeno}
\end{equation}
but rather the average of the four plaquettes lying in the plane
$\mu \nu$, stemming from the point $n$:
\begin{equation}
F_{n, \mu \nu} = \frac{1}{4} \sum_{\mu \nu = \pm} P_{n, \mu \nu} =
\frac{1}{8\mathrm{i}g_0a^2} \sum_{\mu \nu = \pm} (U_{n, \mu \nu}
- U^{+}_{n, \mu \nu}). \label{eq:effesi}
\end{equation}}
\begin{equation}
\Delta S^{f}_{\mathrm{I}} = - \mathrm{i}g_0 a^{4} \sum_{n,\mu \nu}
\frac{r}{4 a} \: \overline{\psi}_{n}
\sigma_{\mu \nu} F_{n, \mu \nu} \psi_{n} \label{eq:impr}.
\end{equation}
This term modifies the standard order $g_0$ quark-gluon Wilson vertex,
\begin{equation}
(V)^{bc}_{\rho} (k,k') = -g_0 (t^A)_{bc} \left[ r \sin
\frac{a(k + k')_{\rho}}{2} + \mathrm{i}\gamma_{\rho}
\cos \frac{a(k + k')_{\rho}}{2} \right] \label{eq:standWil},
\end{equation}
by adding to it the extra piece (``improved vertex'')
\begin{equation}
(V^{\mathrm{I}})^{bc}_{\rho} (k,k') = -g_0 \frac{r}{2} (t^A)_{bc}
\cos \frac{a(k - k')_{\rho}}{2} \sum_{\lambda}
\sigma_{\rho \lambda} \sin a(k - k')_{\lambda} \label{eq:newimpr},
\end{equation}
where $k$ and $k'$ are the momenta of the incoming and the outgoing
quark respectively and $\rho$ is the Lorentz index carried by the
gluon. The quark and gluon propagators, as well as the interaction
between the quark current and an even number of gluons, turn out to be
unmodified. To the order to which we will perform our calculations
we will not need the expression of higher order quark-gluon
vertices.

Besides adding the term (\ref{eq:impr}) to the action, for consistency
\cite{rom1} one also has to rotate all the quark fields appearing in
Green functions according to the rule (see Appendix A for the
definition of the lattice covariant derivatives)
\begin{displaymath}
\psi \longrightarrow \left( 1 - \frac{a r}{2} \stackrel{\displaystyle
\rightarrow }{\FMSlash{D}} \right) \psi
\end{displaymath}
\begin{equation}
\overline{\psi} \longrightarrow \overline{\psi} \left( 1 +
\frac{a r}{2} \stackrel{\displaystyle \leftarrow }{\FMSlash{D}} \right)
\label{eq:rotation}.
\end{equation}
As it has already been noticed in \cite{first}, the transformations
(\ref{eq:rotation}) are the sources of very many algebraic
complications in perturbative calculations.

\section{Structure of the operators}

In this paper we want to study the lattice renormalization of
the rank three quark operator
\begin{equation}
O^q_{\{\mu \nu \tau\}} = \frac{1}{8} \,\overline{\psi} \,
\gamma_{\{ \mu} \! \stackrel{\displaystyle \leftrightarrow}{D}_{\nu}
\stackrel{\displaystyle \leftrightarrow}{D}_{\tau\}} \psi
\label{eq:o3} .
\end{equation}
Its hadronic matrix elements are related to the second moment of the
$x_B$-distribution of quarks inside the hadron.
$O^q_{\{\mu \nu \tau\}}$ is the (parity conserving piece of the)
second operator in the list given in Eqs.~(\ref{eq:opsns}).

We will restrict ourselves to the quenched approximation, and
consequently we will not have to consider the mixing between
the quark operator (\ref{eq:o3}) and the gluon operator
$\sum_{\rho} \mathrm{Tr} \left[ F_{\{\mu \rho} \stackrel{\displaystyle
\leftrightarrow}{D}_{\nu} F_{\rho \tau\}} \right]$.
Actually both the experimental and the numerical determinations
of the $x_B$-distribution of gluons are much harder to obtain
than the corresponding quantities for quarks.

\subsection{General considerations}

The explicit expression of the ``improved'' quark operator, obtained
by taking into account the rotations (\ref{eq:rotation}) on the
fermion fields, is given by
\begin{displaymath}
(O^q_{\{\mu \nu \tau\}})^{\mathrm{I}} =
\end{displaymath}
\begin{displaymath}
\frac{1}{8} \left[
\overline{\psi} \gamma_{\{\mu}
\stackrel{\displaystyle \rightarrow}{D}_{\nu}
\stackrel{\displaystyle \rightarrow}{D}_{\tau\}} \psi
- \overline{\psi} \gamma_{\{\mu}
\stackrel{\displaystyle \leftarrow}{D}_{\nu}
\stackrel{\displaystyle \rightarrow}{D}_{\tau\}} \psi
- \overline{\psi} \gamma_{\{\mu}
\stackrel{\displaystyle \rightarrow}{D}_{\nu}
\stackrel{\displaystyle \leftarrow}{D}_{\tau\}} \psi
+ \overline{\psi} \gamma_{\{\mu}
\stackrel{\displaystyle \leftarrow}{D}_{\nu}
\stackrel{\displaystyle \leftarrow}{D}_{\tau\}} \psi
\right]
\end{displaymath}
\begin{displaymath}
- \frac{a r}{16} \left[
\overline{\psi} \gamma_{\{\mu}
\stackrel{\displaystyle \rightarrow}{D}_{\nu}
\stackrel{\displaystyle \rightarrow}{D}_{\tau\}}
\stackrel{\displaystyle \rightarrow }{\FMSlash{D}} \psi
- \overline{\psi} \gamma_{\{\mu}
\stackrel{\displaystyle \leftarrow}{D}_{\nu}
\stackrel{\displaystyle \rightarrow}{D}_{\tau\}}
\stackrel{\displaystyle \rightarrow }{\FMSlash{D}} \psi
\right.
\end{displaymath}
\begin{displaymath}
- \overline{\psi} \gamma_{\{\mu}
\stackrel{\displaystyle \rightarrow}{D}_{\nu}
\stackrel{\displaystyle \leftarrow}{D}_{\tau\}}
\stackrel{\displaystyle \rightarrow }{\FMSlash{D}} \psi
+ \overline{\psi} \gamma_{\{\mu}
\stackrel{\displaystyle \leftarrow}{D}_{\nu}
\stackrel{\displaystyle \leftarrow}{D}_{\tau\}}
\stackrel{\displaystyle \rightarrow }{\FMSlash{D}} \psi
- \overline{\psi}
\stackrel{\displaystyle \leftarrow }{\FMSlash{D}}
\gamma_{\{\mu}
\stackrel{\displaystyle \rightarrow}{D}_{\nu}
\stackrel{\displaystyle \rightarrow}{D}_{\tau\}} \psi
\end{displaymath}
\begin{displaymath}
\left.
+ \overline{\psi} \stackrel{\displaystyle \leftarrow }{\FMSlash{D}}
\gamma_{\{\mu} \stackrel{\displaystyle \leftarrow}{D}_{\nu}
\stackrel{\displaystyle \rightarrow}{D}_{\tau\}} \psi
+ \overline{\psi} \stackrel{\displaystyle \leftarrow }{\FMSlash{D}}
\gamma_{\{\mu} \stackrel{\displaystyle \rightarrow}{D}_{\nu}
\stackrel{\displaystyle \leftarrow}{D}_{\tau\}} \psi
- \overline{\psi} \stackrel{\displaystyle \leftarrow }{\FMSlash{D}}
\gamma_{\{\mu} \stackrel{\displaystyle \leftarrow}{D}_{\nu}
 \stackrel{\displaystyle \leftarrow}{D}_{\tau\}} \psi
\right]
\end{displaymath}
\begin{displaymath}
- \frac{a^2 r^2}{32} \left[
\overline{\psi} \stackrel{\displaystyle \leftarrow }{\FMSlash{D}}
\gamma_{\{\mu} \stackrel{\displaystyle \rightarrow}{D}_{\nu}
\stackrel{\displaystyle \rightarrow}{D}_{\tau\}}
\stackrel{\displaystyle \rightarrow }{\FMSlash{D}} \psi
- \overline{\psi} \stackrel{\displaystyle \leftarrow }{\FMSlash{D}}
\gamma_{\{\mu} \stackrel{\displaystyle \leftarrow}{D}_{\nu}
\stackrel{\displaystyle \rightarrow}{D}_{\tau\}}
\stackrel{\displaystyle \rightarrow }{\FMSlash{D}} \psi
\right.
\end{displaymath}
\begin{equation}
\left.
- \overline{\psi} \stackrel{\displaystyle \leftarrow }{\FMSlash{D}}
\gamma_{\{\mu} \stackrel{\displaystyle \rightarrow}{D}_{\nu}
\stackrel{\displaystyle \leftarrow}{D}_{\tau\}}
\stackrel{\displaystyle \rightarrow }{\FMSlash{D}} \psi
+ \overline{\psi} \stackrel{\displaystyle \leftarrow }{\FMSlash{D}}
\gamma_{\{\mu} \stackrel{\displaystyle \leftarrow}{D}_{\nu}
 \stackrel{\displaystyle \leftarrow}{D}_{\tau\}}
\stackrel{\displaystyle \rightarrow }{\FMSlash{D}} \psi
\right] \label{eq:o3impr} .
\end{equation}

As can be seen, the three-index case introduces a new feature in the
expression of improved operators, namely the appearance of the
cross-derivative term, $ \overline{\psi}
\stackrel{\displaystyle \rightarrow }{D_{\mu}}
\stackrel{\displaystyle \leftarrow }{D_{\nu}} \psi $,
which has to be defined according to the formula
\begin{displaymath}
\overline{\psi}_n \stackrel{\displaystyle \rightarrow }{D_{\mu}}
\stackrel{\displaystyle \leftarrow }{D_{\nu}} \psi_n =
-\frac{1}{4 a^2} \Bigg[
\overline{\psi}_{n - \nu} U_{n - \nu, \mu} U_{n + \mu - \nu, \nu}
\psi_{n + \mu}
-\overline{\psi}_{n + \nu} U_{n + \nu, \mu} U^{+}_{n + \mu, \nu}
\psi_{n + \mu}
\end{displaymath}
\begin{equation}
-\overline{\psi}_{n - \nu} U^{+}_{n - \mu - \nu, \mu}
U_{n - \mu - \nu, \nu} \psi_{n - \mu}
+\overline{\psi}_{n + \nu} U^{+}_{n - \mu + \nu, \mu}
U^{+}_{n - \mu, \nu} \psi_{n - \mu} \Bigg] .
\end{equation}
The more usual terms
$ \overline{\psi} \stackrel{\displaystyle \rightarrow }{D_{\mu}}
\stackrel{\displaystyle \rightarrow }{D_{\nu}} \psi $,
$ \overline{\psi} \stackrel{\displaystyle \leftarrow }{D_{\mu}}
\stackrel{\displaystyle \leftarrow }{D_{\nu}} \psi $ and
$ \overline{\psi} \stackrel{\displaystyle \leftarrow }{D_{\mu}}
\stackrel{\displaystyle \rightarrow }{D_{\nu}} \psi $
have respectively the expressions
\begin{displaymath}
\overline{\psi}_n \stackrel{\displaystyle \rightarrow }{D_{\mu}}
\stackrel{\displaystyle \rightarrow }{D_{\nu}} \psi_n =
\frac{1}{4 a^2} \overline{\psi}_n \Bigg[
U_{n, \mu} U_{n + \mu, \nu} \psi_{n + \mu + \nu}
-U_{n, \mu} U^{+}_{n + \mu - \nu, \nu} \psi_{n + \mu - \nu}
\end{displaymath}
\begin{equation}
-U^{+}_{n - \mu, \mu} U_{n - \mu, \nu} \psi_{n - \mu + \nu}
+U^{+}_{n - \mu, \mu} U^{+}_{n - \mu - \nu, \nu} \psi_{n - \mu - \nu}
\Bigg]
\end{equation}
\begin{displaymath}
\overline{\psi}_n \stackrel{\displaystyle \leftarrow }{D_{\mu}}
\stackrel{\displaystyle \rightarrow }{D_{\nu}} \psi_n =
- \frac{1}{4 a^2} \Bigg[
\overline{\psi}_{n - \mu} U_{n - \mu, \mu} U_{n, \nu} \psi_{n + \nu}
-\overline{\psi}_{n + \mu} U^{+}_{n, \mu} U_{n, \nu} \psi_{n + \nu}
\end{displaymath}
\begin{equation}
-\overline{\psi}_{n - \mu} U_{n - \mu, \mu} U^{+}_{n - \nu, \nu}
\psi_{n - \nu} +\overline{\psi}_{n + \mu} U^{+}_{n, \mu}
U^{+}_{n - \nu, \nu} \psi_{n - \nu} \Bigg]
\end{equation}
\begin{displaymath}
\overline{\psi}_n \stackrel{\displaystyle \leftarrow }{D_{\mu}}
\stackrel{\displaystyle \leftarrow }{D_{\nu}} \psi_n =
\frac{1}{4 a^2} \Bigg[
\overline{\psi}_{n + \mu + \nu} U^{+}_{n + \nu, \mu} U^{+}_{n, \nu}
-\overline{\psi}_{n - \mu + \nu} U_{n - \mu + \nu, \mu} U^{+}_{n, \nu}
\end{displaymath}
\begin{equation}
-\overline{\psi}_{n + \mu - \nu} U^{+}_{n - \nu, \mu} U_{n - \nu, \nu}
+\overline{\psi}_{n - \mu - \nu} U_{n - \mu - \nu, \mu}
U_{n - \nu, \nu} \Bigg] \psi_n  .
\label{eq:complicat}
\end{equation}
The requirement that fixes the exact form of the cross-derivative
term is that the right-derivative, $\stackrel{\displaystyle
\rightarrow }{D_{\nu}}$, and the corresponding left-derivative,
$\stackrel{\displaystyle \leftarrow }{D_{\nu}}$, should be obtained
one from the other by an integration by parts, that amounts on the
lattice to a shift of one site in the $\nu$ (in this example)
direction.

Using the nearest-neighbor improved action, we want to compute
to 1-loop in the chiral limit the matrix elements
\begin{equation}
M_{\mu \nu \tau} (p) \equiv\langle p | O^q_{\mu \nu \tau} | p\rangle,
\label{eq:amplit}
\end{equation}
where $|p \rangle$ is a one-quark state of momentum $p$ and vanishing
(renormalized) mass. The renormalization constants and mixing
coefficients of $O^q_{\mu \nu \tau}$ can be extracted from
the knowledge of the amplitudes $M_{\mu \nu \tau} (p)$.

\subsection{Operator mixing}

Lorentz indices in Eq.~(\ref{eq:amplit})
have to be chosen carefully by looking at the
transformation properties of the resulting operators with respect
to the hypercubic group, H(4) \cite{mix1,mix2}. This group, which
consists of the discrete proper rotations (no parity
reflections\footnote{In the following we will restrict ourselves
to the transformation properties of the operators under the proper
hypercubic group, as we will only consider unpolarized structure
functions, in which case the $\gamma_5$ contributions drop out from
Eqs.~(\ref{eq:opsns}).}) in four dimensions, is the remnant of the
proper euclidean Lorentz group on a discretized space-time.

In moving from a continuum four-dimensional relativistic theory
to its lattice version, the proper (euclidean) Lorentz invariance
O(4) is broken to the invariance under the proper hypercubic group
H(4). For this reason representations that under the O(4) group are
irreducible become in general reducible under H(4) when the
corresponding operators are written on the lattice. Therefore,
some special care must be exerted to avoid unwanted mixings.

The most dangerous situation occurs if one considers
the operator $O^q_{\mu \nu \tau}$ with $\mu=\nu=\tau$, because it
can mix (and indeed it does) with the lower-dimensional operator
\begin{equation}
\overline{\psi} \:\gamma_{\mu} \:\psi,
\end{equation}
with a power divergent ($\sim 1 / a^2$) mixing coefficient.
Any simulation carried out with an operator with three equal indices
will require a delicate non-perturbative subtraction of the
lower-dimensional operator. This ends up in very large statistical
errors in the results of these simulations.

Another possibility is to take $\mu\neq\nu\neq\tau (\neq\mu)$.
In this way one selects an irreducible representation of H(4),
and no mixing is anymore allowed. We will take as a typical component
of the multiplet of operators transforming according to
this representation the operator $O^q_{\{123\}}$. The drawback
of choosing $\mu\neq\nu\neq\tau (\neq\mu)$
in a Monte Carlo simulation is that two components
of the momentum of the hadron are bound to be different from zero
and from each other. This can lead to significant
systematical errors coming from the increased sensitivity to the
granularity of the lattice.

A third possible choice (and the best one) is $\mu=\nu\neq\tau$.
In this case only one component of the hadron momentum has to be
taken different from zero.
Martinelli and Sachrajda have shown that power divergent subtractions
can be avoided by using the particular combination \cite{Mart1,Mart2}
\begin{equation}
O^q_{\mathrm{DIS}} \equiv
O^q_{\{411\}} - \frac{1}{2} (O^q_{\{422\}} + O^q_{\{433\}}),
\label{eq:otto}
\end{equation}
which turns out to belong to an irreducible representation of H(4).
Using the standard Wilson action, they estimated the renormalization
constant of this operator in the hypothesis of tadpole dominance
of lattice perturbation theory.

Unfortunately, the fact that the operator (\ref{eq:otto})
belongs to an irreducible representation of H(4) is not
enough for it to be multiplicatively renormalizable,
because in the decomposition of a rank three tensor into
irreducible representations, the representation to which the
operator (\ref{eq:otto}) belongs appears more than once.
As a consequence, mixing can occur among operators belonging to
these equivalent representations.

In the next subsection we will present the complete solution to the
problem of decomposing in irreducible H(4) representations
the particular tensor products of representations that arise
when one considers the (parity conserving pieces of the) DIS quark
operators of rank 3, 4 and 5 appearing in Eqs.~(\ref{eq:opsns}).

\subsection{Lattice transformation properties of DIS quark operators}

The proper hypercubic group H(4) has 192 elements. Its 13 irreducible
representations are listed in Table 1, where an extra subscript
is used to label different representations with the same dimension.
For example, the representation labelled as $\mbox{\bf 1}_1$ is the
identity, while the representation labelled as $\mbox{\bf 1}_2$ is
the totally antisymmetric one. In Table 1 we also give the
corresponding O(4) representation in SU(2) $\otimes$ SU(2) notation,
when it exists. We recall that the representations $\mbox{\bf 3}_5$,
$\mbox{\bf 3}_6$ and $\mbox{\bf 4}_2$ can be obtained as tensor
product of $\mbox{\bf 1}_2$ with $\mbox{\bf 3}_3$, $\mbox{\bf 3}_4$
and $\mbox{\bf 4}_1$ respectively. With an abuse of notation, in the
literature they are often denoted respectively as
$\overline{(1,0)}$, $\overline{(0,1)}$ and
$\overline{(\half,\half)}$.

\begin{table}
\caption{The irreducible representations of the hypercubic group
(the corresponding O(4) representations are also indicated)}
\begin{tabular}{ccccccccccccc}
\hline $\mbox{\bf 1}_1$ & $\mbox{\bf 1}_2$ & $\mbox{\bf 2}$
& $\mbox{\bf 3}_1$ & $\mbox{\bf 3}_2$ & $\mbox{\bf 3}_3$
& $\mbox{\bf 3}_4$ & $\mbox{\bf 3}_5$ & $\mbox{\bf 3}_6$
& $\mbox{\bf 4}_1$ & $\mbox{\bf 4}_2$ & $\mbox{\bf 6}$
& $\mbox{\bf 8}$ \\
$(0,0)$ & $-$ & $-$ & $-$ & $-$ & $(1,0)$ & $(0,1)$
& $-$ & $-$ & $(\half,\half)$ & $-$ & $-$ & $-$ \\
\hline
\end{tabular}
\end{table}

We are interested in the decomposition of tensor products such as
$\mbox{\bf 4}_1 \otimes \mbox{\bf 4}_1 \otimes \cdots \otimes
\mbox{\bf 4}_1$ to which all the operators listed in
Eqs.~(\ref{eq:opsns}) belong. According to Table 1, the irreducible
$\mbox{\bf 4}_1$ representation of H(4) corresponds to the irreducible
vector representation $(\half,\half)$ of O(4).

The case of the rank three operator is easily worked out remembering
the corresponding O(4) decomposition formula
\begin{equation}
(\frac{1}{2}, \frac{1}{2}) \otimes
(\frac{1}{2}, \frac{1}{2}) \otimes
(\frac{1}{2}, \frac{1}{2}) =
4 \cdot (\frac{1}{2}, \frac{1}{2})
+ 2 \cdot (\frac{1}{2}, \frac{3}{2})
+ 2 \cdot (\frac{3}{2}, \frac{1}{2})
+ (\frac{3}{2}, \frac{3}{2}) \label{eq:444c}.
\end{equation}
One finds for H(4)
\begin{equation}
\mbox{\bf 4}_1 \otimes \mbox{\bf 4}_1 \otimes \mbox{\bf 4}_1 = 5 \cdot
\mbox{\bf 4}_1 + \mbox{\bf 4}_2 + 5 \cdot \mbox{\bf 8} \label{eq:444}.
\end{equation}
These representations can be organized according to the patterns and
number of indices having the same value, as shown in Table 2.
In the first column we give a typical representative for each type
of index pattern.

\begin{table}
\caption{Multiplet structure for $\mbox{\bf 4}_1 \otimes
\mbox{\bf 4}_1 \otimes \mbox{\bf 4}_1$}
\begin{tabular}{cl}
\hline typical index pattern & multiplets \\
\hline 123 & $ \mbox{\bf 4}_1 + \mbox{\bf 4}_2 + \mbox{\bf 8} +
\mbox{\bf 8} $ \\
\hline 112 & $ 3 \cdot ( \mbox{\bf 4}_1 + \mbox{\bf 8} ) $ \\
\hline 111 & $\mbox{\bf 4}_1$ \\
\hline
\end{tabular}
\end{table}

It is straightforward although tedious to extend the above results
to the more complicated cases of operators of rank four and five.
The H(4) decomposition of the 256 components of the
lattice rank four operator is given by
\begin{displaymath}
\mbox{\bf 4}_1 \otimes \mbox{\bf 4}_1 \otimes \mbox{\bf 4}_1 \otimes
\mbox{\bf 4}_1 =
5 \cdot \mbox{\bf 1}_1 + \mbox{\bf 1}_2 + 5 \cdot \mbox{\bf 2}
+ 10 \cdot \mbox{\bf 3}_1 + 6 \cdot \mbox{\bf 3}_2
\end{displaymath}
\begin{equation}
+ 10 \cdot \mbox{\bf 3}_3 + 10 \cdot \mbox{\bf 3}_4
+ 6 \cdot \mbox{\bf 3}_5 + 6 \cdot \mbox{\bf 3}_6
+ 16 \cdot \mbox{\bf 6} \label{eq:4444}.
\end{equation}
The decomposition results in 75 multiplets belonging to ten
different types of representations, which again can be organized
according to the patterns and number of indices having the same
value, as shown in Table 3.

\begin{table}
\caption{Multiplet structure for $\mbox{\bf 4}_1 \otimes
\mbox{\bf 4}_1 \otimes \mbox{\bf 4}_1 \otimes \mbox{\bf 4}_1$}
\begin{tabular}{cl}
\hline typical index pattern & multiplets \\
\hline 1234 & $ \mbox{\bf 1}_1 + \mbox{\bf 1}_2 + \mbox{\bf 2}
+ \mbox{\bf 2} + \mbox{\bf 3}_1 + \mbox{\bf 3}_1 + \mbox{\bf 3}_1
+ \mbox{\bf 3}_2 + \mbox{\bf 3}_2 + \mbox{\bf 3}_2 $ \\
\hline 1123 & $ 6 \cdot ( \mbox{\bf 3}_3 + \mbox{\bf 3}_4
+ \mbox{\bf 3}_5 + \mbox{\bf 3}_6 + \mbox{\bf  6} + \mbox{\bf 6})$\\
\hline 1122 & $ 3 \cdot ( \mbox{\bf 1}_1 + \mbox{\bf 2}
+ \mbox{\bf 3}_1 + \mbox{\bf 3}_1 + \mbox{\bf 3}_2 ) $ \\
\hline 1112 & $ 4 \cdot ( \mbox{\bf 3}_3 + \mbox{\bf 3}_4
+ \mbox{\bf 6} ) $ \\
\hline 1111 & $ \mbox{\bf 1}_1 + \mbox{\bf 3}_1 $ \\
\hline
\end{tabular}
\end{table}

Finally, the 1024 components of the lattice rank five operator
decompose in 171 multiplets of only three different kinds,
\begin{equation}
\mbox{\bf 4}_1 \otimes \mbox{\bf 4}_1 \otimes \mbox{\bf 4}_1 \otimes
\mbox{\bf 4}_1 \otimes \mbox{\bf 4}_1 = 51 \cdot \mbox{\bf 4}_1
+ 35 \cdot \mbox{\bf 4}_2 + 85 \cdot \mbox{\bf 8} \label{eq:44444},
\end{equation}
and their classification is reported in Table 4.

\begin{table}
\caption{Multiplet structure for $\mbox{\bf 4}_1 \otimes
\mbox{\bf 4}_1 \otimes \mbox{\bf 4}_1 \otimes \mbox{\bf 4}_1
\otimes \mbox{\bf 4}_1$}
\begin{tabular}{cl}
\hline typical index pattern & multiplets \\
\hline 11234 & $ 10 \cdot ( \mbox{\bf 4}_1 + \mbox{\bf 4}_2
+ \mbox{\bf 8} + \mbox{\bf 8}) $ \\
\hline 11223 & $ 15 \cdot ( \mbox{\bf 4}_1 + \mbox{\bf 4}_2
+ \mbox{\bf 8} + \mbox{\bf 8}) $ \\
\hline 11222 & $ 10 \cdot ( \mbox{\bf 4}_1 + \mbox{\bf 8}) $ \\
\hline 11123 & $ 10 \cdot ( \mbox{\bf 4}_1 + \mbox{\bf 4}_2
+ \mbox{\bf 8} + \mbox{\bf 8}) $ \\
\hline 11112 & $ 5 \cdot ( \mbox{\bf 4}_1 + \mbox{\bf 8}) $ \\
\hline 11111 & $ \mbox{\bf 4}_1 $ \\
\hline
\end{tabular}
\end{table}

\subsection{Choice of the rank three operators}

It is easily seen that the operator (\ref{eq:otto}) belongs to the
eight-dimensional representation of H(4). Furthermore, we see from
Eq.~(\ref{eq:444}) that when one decomposes $O^q_{\mu \nu \tau}$ in
H(4) multiplets there arise five eight-dimensional representations.
The crucial point here is that the hypercubic group H(4)
possesses only one single eight-dimensional irreducible representation
(see Table 1). Therefore all the five $\mbox{\bf 8}$'s
that come out in the decomposition of the rank three operators into
H(4) multiplets can in principle mix among themselves.
This is at variance with what happens in the case of the rank
two operator, where the operator $O^q_{\{41\}}$ is multiplicatively
renormalizable because it belongs to the irreducible
six-dimensional representation, $\mbox{\bf 6}$, which appears only
once in the decomposition $\mbox{\bf 4}_1 \otimes \mbox{\bf 4}_1
 = \mbox{\bf 1}_1 + \mbox{\bf 3}_1 + \mbox{\bf 3}_3
 + \mbox{\bf 3}_4 + \mbox{\bf 6}$.

A natural question to ask is what are the equivalent representations
that can really mix among themselves in the decompositions
(\ref{eq:444}), (\ref{eq:4444}) and (\ref{eq:44444}). For a general
answer see Ref.~\cite{z2}. For our purposes it is enough to note
here that radiative corrections can not only induce
any permutation of the values of the indices, but can also
make pairs of equal indices flip their common value.
As an example, from $O_{112}$ radiative corrections may generate
$O_{211}$, $O_{442}$, $O_{323}$ and the like, but not $O_{123}$
nor $O_{114}$. Similarly $O_{111}$ can mix with $O_{133}$, $O_{414}$
and so on. But from $O_{123}$ only $O_{213}$ and permutations can
arise, and not operators like $O_{122}$ or $O_{441}$.

For the case of the operator (\ref{eq:otto}) we thus expect mixing
only among the three eight-dimensional representations that
can be constructed by suitably arranging the operator components
that have two and only two equal indices\footnote{No mixing can occur
with the operator $O^q_{\{41\}}$, because as we said it belongs
to the representation $\mbox{\bf 6}$, which does not appear
in the decomposition (\ref{eq:444}).}. To deal with the mixing
of these three copies of the eight-dimensional representation it is
more convenient to consider the following three eight-dimensional
representations\footnote{Apparently the number of components of each
multiplet is 12 instead of 8. But in fact there are 4 constraints,
because the sum of the four components of each line
in (\ref{eq:tre}) is zero.}:
\begin{eqnarray}
i) && O^q_{411} - \frac{1}{2} (O^q_{422} + O^q_{433})
,~~O^q_{422} - \frac{1}{2} (O^q_{411} + O^q_{433}),~~\ldots
\nonumber \\
&& O^q_{311} - \frac{1}{2} (O^q_{322} + O^q_{344})
,~~O^q_{322} - \frac{1}{2} (O^q_{311} + O^q_{344}),~~\ldots
\nonumber \\
&&~~\vdots \nonumber \\
ii) && O^q_{141} - \frac{1}{2} (O^q_{242} + O^q_{343}),~~\ldots
\nonumber \\
&&~~\vdots \nonumber \\
iii) && O^q_{114} - \frac{1}{2} (O^q_{224} + O^q_{334}),~~\ldots
\label{eq:tre} \\
&&~~\vdots \nonumber
\end{eqnarray}
where neither symmetrization nor antisymmetrization of the indices
is to be understood. The point here is that the operator
(\ref{eq:otto}), which is the symmetric combination of the three
kinds of operators in Eqs.~(\ref{eq:tre}), actually mixes with the
two other independent linear combinations of them that do not
possess any special symmetry. In other words, the interaction
may change the symmetry properties of these operators.

In the continuum the transformation properties for the rank three
operators are much simpler. In fact, the symmetric combination
(\ref{eq:otto}) belongs to the sixteen-dimensional
$(\threehalf,\threehalf)$ irreducible representation of O(4),
which only appears once in the decomposition (\ref{eq:444c}),
and consequently the operator (\ref{eq:otto}) is multiplicatively
renormalizable. It follows that in the continuum, unlike
what happens on the lattice, the two four-component multiplets
of operators $(O^q_{\{111\}},O^q_{\{222\}},\ldots)$ and
$(O^q_{\{123\}},O^q_{\{234\}},\ldots)$, which are part of this
representation, have the same renormalization constants as the
operator (\ref{eq:otto}). We report in Table 5 the value of this
common renormalization constant in the $\overline{MS}$ scheme.
Vice versa no mixing occurs between the operator (\ref{eq:otto})
and the two other multiplets of operators with which it mixes
on the lattice, because they can be arranged in the two irreducible
O(4) representations $(\half,\threehalf)$ and $(\threehalf,\half)$.

As for the operator $O_{\{123\}}$, on the lattice it belongs to
the $\mbox{\bf 4}_2$ representation. This representation appears
only once in the product $\mbox{\bf 4}_1 \otimes \mbox{\bf 4}_1
\otimes \mbox{\bf 4}_1$, implying that the operator $O_{\{123\}}$
is multiplicatively renormalizable.

In this work we have computed the renormalization constant of
$O_{\{123\}}$ and the full mixing matrix of the operators
(\ref{eq:tre}). We have also verified that, as expected,
on the lattice the traceless operator $O_{\{111\}}$ is not
multiplicatively renormalizable, and that it indeed mixes with
the three operators belonging to the $\mbox{\bf 4}_1$
representations with two equal indices, as indicated in
the second row of Table 2.

\subsection{On the feasibility of higher moments computations}

We conclude this section with a few observations on the operators
of rank four and five.

In the case of the rank four operators, we see from Table 3 that
there are two operators that are multiplicatively
renormalizable. As in the rank three case, they have all
indices different. One of these representations is the totally
symmetric $\mbox{\bf 1}_1$ representation, and the other is the
completely antisymmetric $\mbox{\bf 1}_2$ representation.
The $\mbox{\bf 1}_2$ representation is unique in the decomposition
(\ref{eq:4444}), while the $\mbox{\bf 1}_1$ appears five times,
grouped in three different patterns. But even in this case no
mixing can occur, because if one starts with the pattern of
indices 1234, radiative corrections can never lead to the other
two kinds of patterns which the other four $\mbox{\bf 1}_1$
representations possess, namely 1122 and 1111. For any other
choice of indices mixing will be unavoidable, because all
the other representations are present more than once in the
decomposition (\ref{eq:4444}), and at the same time they have
patterns of indices that can be turned one into the other
by the interaction.

The use of the rank four operator with all indices different
in Monte Carlo simulations would require three components of
the momentum of the hadron to be different from zero,
and different among themselves.

For the rank five operators, the situation is of course even worse.
In fact there are no representations that are present only once
in the decomposition (\ref{eq:44444}). Furthermore, we see from
Table 4 that within a single pattern of indices
(i.e. within a single row) a representation is never
present only once, except for 11111, which is a $\mbox{\bf 4}_1$.
In this case, however, radiative corrections can induce it
to mix with the ten $\mbox{\bf 4}_1$ representations
with index pattern 11222 and with the fifteen $\mbox{\bf 4}_1$
representations with index pattern 11223. Therefore for rank
five operators mixing is unavoidable.

\section{Renormalization conditions}

In the tree approximation the amputated non-vanishing matrix
elements of the operators $O^q_{\mu \nu \tau}$ that we consider
in this work are given by
\begin{equation}
\langle p| O^q_{\{123\}} (a) |p \rangle
\Big|^{\mathrm{tree}}_{\mathrm{amp}}
= - \frac{1}{2} \gamma_{\{1} p_2 p_{3\}}
\label{eq:tree}
\end{equation}
\begin{displaymath}
\langle p| O^q_{411} - \frac{1}{2} (O^q_{422} + O^q_{433}) (a)
|p \rangle \Big|^{\mathrm{tree}}_{\mathrm{amp}}
= - \frac{1}{2}
\left(\gamma_4 p_1^2
-\frac{1}{2} (\gamma_4 p_2^2 + \gamma_4 p_3^2 ) \right)
\end{displaymath}
\begin{displaymath}
\langle p| O^q_{141} - \frac{1}{2} (O^q_{242} + O^q_{343}) (a)
|p \rangle \Big|^{\mathrm{tree}}_{\mathrm{amp}}
= - \frac{1}{2}
\left(\gamma_1 p_1 p_4
-\frac{1}{2} (\gamma_2 p_2 p_4 + \gamma_3 p_3 p_4 ) \right)
\end{displaymath}
\begin{displaymath}
\langle p| O^q_{114} - \frac{1}{2} (O^q_{224} + O^q_{334}) (a)
|p \rangle \Big|^{\mathrm{tree}}_{\mathrm{amp}}
= - \frac{1}{2}
\left(\gamma_1 p_1 p_4
-\frac{1}{2} (\gamma_2 p_2 p_4 + \gamma_3 p_3 p_4 )  \right).
\end{displaymath}

The renormalization conditions that connect the bare operators on the
lattice to the finite operators renormalized at a scale $\mu$ must be
the same as the ones imposed in the continuum.
We define finite renormalized operators by requiring their two quark
amputated matrix elements computed at $p^2= \mu^2$ to be equal to the
corresponding tree level amplitudes. The renormalization constant
$Z_{\{123\}}$ for the multiplicatively renormalizable operator
$O^q_{\{123\}}$ is thus fixed in perturbation theory by simply
imposing the renormalization condition:
\begin{equation}
\langle p| \widehat{O}^q_{\{123\}} (\mu) |p \rangle
\Big|_{p^2 = \mu^2} =
\langle p| O^q_{\{123\}} (a) |p\rangle
\Big|^{\mathrm{tree}}_{p^2 = \mu^2},
\label{eq:condit}
\end{equation}
where $Z_{\{123\}}$ is given by
\begin{equation}
\widehat{O}^q_{\{123\}} (\mu) = Z_{\{123\}} (\mu a)
O^q_{\{123\}} (a) \label{eq:condit2}.
\end{equation}

The anomalous dimension of the multiplicatively renormalizable
operator $O^q_{\{123\}}$ is $25/3$. For $\mu a \ll 1$, the
renormalization constant of $O^q_{\{123\}}$ can be written in the form
\begin{equation}
Z_{\{123\}} (\mu a) =  1 - {\displaystyle\frac{\alpha_S}{4 \pi}} C_F
\left( {\displaystyle\frac{25}{3}} \log \mu a + B_{\{123\}} \right) ,
\end{equation}
where $\alpha_S=g_0^2/4 \pi$ and $C_F = (N_c^2-1)/2N_c$ is the
quadratic Casimir of the fundamental representation of SU($N_c$);
in QCD $N_c = 3$ and $C_F=4/3$.
The constant $B_{\{123\}}$ is all we have to know to fix $Z_{\{123\}}$
and is given in Sect.~5 for different values of the
Wilson parameter $r$.

For the case in which two of the indices are equal, we see that
two of the three possible combinations give the same tree level result
(see Eqs.~(\ref{eq:tree})). Therefore we can reduce the whole problem
of mixing to the mixing of the two operators
\begin{equation}
O_A \equiv O^q_{411} - \frac{1}{2} (O^q_{422} + O^q_{433})
\end{equation}
and
\begin{equation}
O_B \equiv O^q_{141} + O^q_{114} - \frac{1}{2} (O^q_{242} + O^q_{224}
+ O^q_{343} + O^q_{334}).
\end{equation}
If we set
\begin{displaymath}
\widehat{O}_{A} = Z_{AA} O_{A} + Z_{AB} O_{B}
\end{displaymath}
\begin{equation}
\widehat{O}_{B} = Z_{BA} O_{A} + Z_{BB} O_{B} ,
\end{equation}
the mixing coefficients will be extracted from
the 1-loop calculations according to the formulae
\begin{displaymath}
\langle p| O_A |p \rangle \Big|^{\mathrm{1-loop}}_{p^2 = \mu^2} =
- \frac{1}{2} \left( Z_{AA}^{-1} \left(\gamma_4 p_1^2
-\frac{1}{2} (\gamma_4 p_2^2 + \gamma_4 p_3^2 )\right) \right.
\end{displaymath}
\begin{displaymath}
\left.~~~~+ 2 Z_{AB}^{-1} \left(\gamma_1 p_1 p_4 -\frac{1}{2}
(\gamma_2 p_2 p_4 + \gamma_3 p_3 p_4 )\right) \right)
\end{displaymath}
\begin{displaymath}
\langle p| O_B |p \rangle \Big|^{\mathrm{1-loop}}_{p^2 = \mu^2}
= - \frac{1}{2} \left( Z_{BA}^{-1} \left(\gamma_4 p_1^2
-\frac{1}{2} (\gamma_4 p_2^2 + \gamma_4 p_3^2 ) \right) \right.
\end{displaymath}
\begin{equation}
\left.~~~~+ 2 Z_{BB}^{-1} \left(\gamma_1 p_1 p_4 -\frac{1}{2}
(\gamma_2 p_2 p_4 + \gamma_3 p_3 p_4 )\right) \right) \label{eq:zm1}.
\end{equation}
Taking into account the form of the matrix of the anomalous
dimensions of $O_A$ and $O_B$ (see Table 11), we can write
\begin{displaymath}
Z_{AA} (\mu a) =  1 - {\displaystyle\frac{\alpha_S}{4 \pi}} C_F
\left( {\displaystyle\frac{13}{3}} \log \mu a + B_{AA} \right)
\end{displaymath}
\begin{displaymath}
Z_{AB} (\mu a) =  - {\displaystyle\frac{\alpha_S}{4 \pi}} C_F
\left( 2 \: \log \mu a + B_{AB} \right)
\end{displaymath}
\begin{displaymath}
Z_{BA} (\mu a) =  - {\displaystyle\frac{\alpha_S}{4 \pi}} C_F
\left( 4 \: \log \mu a + B_{BA} \right)
\end{displaymath}
\begin{equation}
Z_{BB} (\mu a) =  1 - {\displaystyle\frac{\alpha_S}{4 \pi}} C_F
\left( {\displaystyle\frac{19}{3}} \log \mu a + B_{BB} \right) ,
\end{equation}
where the values of the constants $B$ can be found in Sect.~5.

For DIS the physically relevant operator is the symmetric combination
$O^q_{\mathrm{DIS}}$ (Eq.~(\ref{eq:otto})), which is simply given by
\begin{equation}
O^q_{\mathrm{DIS}} = \frac{1}{3} \left( O_A + O_B \right) .
\end{equation}
For it we have
\begin{eqnarray}
\widehat{O}^q_{\mathrm{DIS}} (\mu) & = & \frac{1}{3}
\left( \widehat{O}_A (\mu) + \widehat{O}_B (\mu) \right) \nonumber \\
& = & \frac{1}{3} \left( Z_{AA} + Z_{BA} \right) O_A +
\frac{1}{3} \left( Z_{AB} + Z_{BB} \right) O_B .
\end{eqnarray}

\section{Some aspects of the computation}

The computation of the renormalization constant of the operator
$O_{\{123\}}$ has been crosschecked by comparing the results
obtained from two completely independent codes, written
to automatically carry out all the extremely complicated algebraic
manipulations necessary in the calculation \cite{aihenp}.
In fact, contrary to the case of the rank two operator, no complete
hand checks are possible here\footnote{We have completed the
calculations by hand only for the non-improved diagrams; the
calculations of the improved ones is hard drudgery.}: the operator
of rank three contains one more covariant derivative,
and all propagators, vertices and operators have to
be expanded to second order in $a$. This results in impossibly
long algebraic expressions. In the case of the operator with two
equal indices we have checked by hand the results only for the
simplest Feynman diagrams.

We started our computations by employing
a Schoonschip code \cite{phd} which was obtained as an
evolution of the one we used in Ref.~\cite{first}.
At a later stage we decided to use the more flexible
Form language, and we developed a completely new set of
routines to deal with Dirac $\gamma$-matrices.

The most innovative feature of the new Form program is a routine
that reduces the products of $\gamma$-matrices in a much more
efficient way than before. As a byproduct, the treatment of even
products of sines\footnote{Odd number of sines integrates to zero.}
can be postponed, and performed after the completion
of the Dirac algebra; in this way, we happily no longer need
to evaluate the daring expressions involving six and eight sines
that gave rise to a plethora of terms (see Tables C.2 and C.3 in
Ref.~\cite{first}) in the old Schoonschip programs. One discovers
that what is needed in this new way of proceeding is at most the
peculiar combination of six sines of the kind
\begin{displaymath}
{\cal I}(\mu_1,\mu_2,\mu_3,\mu_4) =
\int\! d^{4} k \, f(\cos k, \sum_{\lambda} \sin^2 k_{\lambda})
\sin k_{\mu_1} \sin k_{\mu_2} \sin k_{\mu_3}
\sin^3 k_{\mu_4} =
\end{displaymath}
\begin{displaymath}
\int\! d^{4} k \, f(\cos k, \sum_{\lambda} \sin^2 k_{\lambda})
\sin^2 k_{\mu_1} \sin^4 k_{\mu_3}
\: \delta_{\mu_1 \mu_2} \: \delta_{\mu_3 \mu_4}
\end{displaymath}
\begin{displaymath}
+\int\! d^{4} k \, f(\cos k, \sum_{\lambda} \sin^2 k_{\lambda})
\sin^2 k_{\mu_1} \sin^4 k_{\mu_2}
\: \delta_{\mu_1 \mu_3} \: \delta_{\mu_2 \mu_4}
\end{displaymath}
\begin{displaymath}
+\int\! d^{4} k \, f(\cos k, \sum_{\lambda} \sin^2 k_{\lambda})
\sin^4 k_{\mu_1} \sin^2 k_{\mu_2}
\: \delta_{\mu_1 \mu_4} \: \delta_{\mu_2 \mu_3}
\end{displaymath}
\begin{equation}
- 2 \cdot \int\! d^{4} k\, f(\cos k, \sum_{\lambda} \sin^2 k_{\lambda})
\sin^6 k_{\mu_1}
\: \delta_{\mu_1 \mu_2 \mu_3 \mu_4} \label{eq:expsines},
\end{equation}
where $\delta_{\mu_1 \mu_2 \mu_3 \mu_4}$ is non-zero only if
all the indices are equal. The structure and the coefficients of
Eq.~(\ref{eq:expsines}) are the same as those that appear in
the similar formula for the case of four sines (see Table C.1
in Ref.~\cite{first}), with a sine raised to the third power
here playing the role of a plain sine there.

The Form code described above is the one that has been carefully
optimized and used more extensively in our computations. Meanwhile,
another Form code, completely independent from it, has been developed
by one of us \cite{beccarini0}. Summarizing, three sets of routines
have grown up from our efforts, and we have checked for each code
separately that the rather complicated Dirac algebra (we have products
of up to seven gamma matrices) is carried out correctly.
The analytical results obtained with the modified old Schoonschip
program and with the two newly developed Form codes have been compared,
and they turn out to be in perfect agreement. This we regard
as a significative check of our calculations.

The numerical integration has also been performed using various
different Fortran programs, that all gave the same results within
a 1\% relative error. For the numerical integration, and for the
treatment of the logarithmic divergences, we have used the same
strategy as in Ref.~\cite{first} and once more we refer the reader
to that paper for details.

A big problem we have encountered was the limitation on the working
memory allowed by Schoonschip and Form on the different machines
we have used. This is of primary relevance in the first stages of the
computation, when all vertices, propagators etc. have to be
expanded up to second order in the lattice spacing $a$.
When dealing with the rank three case, one encounters several
products of up to ten trigonometric functions, each one to be
expanded up to order $a^2$. Every product at first will thus give
rise to something like $10^4$ monomials. A large part of them
does not contribute to the final expression, to the order
in $a$ we are interested in, and has to be killed at the
earliest possible stage of the calculation.
A careless way of programming may lead to an increase
of CPU times of several orders of magnitude, or even to an
abrupt stop of the elaboration for lack of memory.

In going to the new Form programs, and thanks to the evolution of
the various routines, the CPU time needed to perform the whole
analytic calculation was drastically cut down. The CPU time required
by the old Schoonschip code running on a Sun 3 workstation
ranged from a few minutes up to more than four hours for the most
complicated diagrams. The Form program was first running on VAX-VMS
machines and then on a HP-UX 9000/735 machine. In its present form
the computation of a generic diagram only takes from about ten
seconds up to a few minutes.

We conclude this section with a discussion on an important physical
point that needs some clarification. This is the role of the
terms $a^2 p^2$ that appear in what we may call the
effective ``Feynman propagator'', $P(ak,ap;q)$, of a diagram,
i.e. the overall denominator that arises after the introduction
of the Feynman parameters. Generally it will have the form
\begin{equation}
P(ak,ap;q) = \Big[ g(ak) + ap_{\mu} h_{\mu}(ak)
+ \alpha (1 - \alpha) a^2p^2 +a^2 p_{\mu} p_{\nu}
f_{\mu \nu}(ak) \Big]^q,
\end{equation}
with $q \ge 2$, where the functions $h_{\mu}(ak)$ and
$f_{\mu \nu}(ak)$ have the following small $ak$ behavior:
\begin{displaymath}
h_{\mu}(ak) \stackrel{ak \rightarrow 0}{\sim} (ak)^3
\end{displaymath}
\begin{equation}
f_{\mu \nu}(ak) \stackrel{ak \rightarrow 0}{\sim} (ak)^2 .
\end{equation}
We see that the terms $a^2p^2$ enter either multiplied by a function
of $ak$ vanishing as $ak \rightarrow 0$ (e.g. $\sin^2 ak,~1
- \cos ak$ etc.) or by a constant independent of $a$
(i.e. $\alpha (1 - \alpha) a^2p^2$).
They should be treated differently according to whether the diagram
has already the required $p_\mu p_\nu$ tensor structure in front or
whether this structure will emerge after expanding the Feynman
denominator itself. For dimensional reasons in the first case the
diagram will be logarithmically divergent. Terms such as $ap_{\mu}
h_{\mu}(ak)$ and $a^2p_{\mu} p_{\nu} f_{\mu \nu}(ak)$ are to be
set to zero, while the term $\alpha (1 - \alpha) a^2p^2$ must be kept
in the denominator and will act as a regulator for the lattice infrared
singularity at $ak=0$. In the limit $ap \ll 1$ it will give rise to
the expected logarithmic behavior, $\log ap$. In the second case the
integral is finite because a function of $ak$, vanishing as $(ak)^2$
in the limit $ak \rightarrow 0$, will dimensionally replace
in the numerator the two missing momenta present in the previous case.
In this situation the contribution of the diagram to the sought tensor
structure is obtained by expanding the Feynman propagator in powers
of $a^2p_{\mu} p_{\nu} f_{\mu \nu}(ak)$.
Each term in this expansion is finite, thanks to the
smoothing out provided by the function $f_{\mu \nu}(ak)$,
and one is allowed to set to zero the $\alpha (1 - \alpha) a^2p^2$
term. Since we are interested in the terms that have at most
two powers of $p$ in front, we must look in the
expansion only to the terms that are linear in
$a^2p_{\mu} p_{\nu} f_{\mu \nu}(ak)$.

Similarly $P(ak,ap;q)$ must be expanded in powers of
$ap_{\mu} h_{\mu}(ak)$ up to second order. Linear and quadratic
terms must be kept in the expansion to combine with the possible
$p$-dependence present in the numerator and
reconstruct the required $p_{\mu} p_{\nu}$ tensor structure.
The coefficients of this expansion are finite loop momentum integrals
in which all $a^2p^2$ terms can be safely set to zero.

\section{Results}

\begin{table}
\caption{Values of the constant $B_{(\threehalf,\threehalf)}$ in the
$\overline{MS}$ regularization scheme}
\begin{tabular}{crr}
\hline & sails & -62$/$9 \\
$B^{\overline{MS}}_{(\threehalf,\threehalf)}$ & vertex & 4$/$9 \\
& self-energy & -1 \\
& total & - 67$/$9 \\
\hline
\end{tabular}
\end{table}

\begin{table}
\caption{Values of $B_{\{123\}}$}
\begin{tabular}{ccrrrr}
\hline  & {\sl r} & Wilson & $O(a)$ impr. & $O(a^2)$ impr. & total \\
\hline  & {\sl 0.2} & -4.172 & -1.611 & -0.022 & -5.805 \\
  & {\sl 0.4}& -5.368 & -3.683 & -0.106 & -9.157 \\
  {\bf SAILS} & {\sl 0.6}& -6.309 & -5.033 & -0.271 & -11.612 \\
  & {\sl 0.8} & -7.032 & -5.838 & -0.520 & -13.389 \\
  & {\sl 1.0}& -7.607 & -6.313 & -0.854 & -14.773 \\
\hline  & {\sl 0.2} & 0.387 & -0.060 & 0.012 & 0.339 \\
  & {\sl 0.4} & 0.574 & -0.215 & 0.064 & 0.423 \\
  {\bf VERTEX} & {\sl 0.6} & 0.804 & -0.398 & 0.162 & 0.568 \\
  & {\sl 0.8} & 1.021 & -0.572 & 0.307 &  0.756 \\
  & {\sl 1.0} & 1.216 & -0.731 & 0.500 & 0.984 \\
\hline  & {\sl 0.2} & -6.102 & -0.611 & 0 & -6.713 \\
  & {\sl 0.4} & -4.326 & -1.513 & 0 & -5.839 \\
  {\bf $\half$ SELF-ENERGY} & {\sl 0.6} & -2.762 & -2.311 & 0 & -5.073
  \\
  & {\sl 0.8} & -1.465 & -3.013 & 0 & -4.479 \\
  & {\sl 1.0} & -0.381 & -3.646 & 0 & -4.027 \\
\hline  & {\sl 0.2} & -12.233 & 0 & 0.009 & -12.224 \\
  {\bf operator TADPOLE} & {\sl 0.4} & -12.233 & 0 & 0.037 & -12.196 \\
  $+$ & {\sl 0.6} & -12.233 & 0 & 0.084 & -12.149 \\
  {\bf $\half$ leg TADPOLE} & {\sl 0.8} & -12.233 & 0 & 0.150
  & -12.083 \\
  & {\sl 1.0} & -12.233 & 0 & 0.234 & -11.999 \\
\hline \hline & {\sl 0.2} & -22.121 & -2.282 & -0.000 & -24.403 \\
  & {\sl 0.4} & -21.353 & -5.411 & -0.005 & -26.769 \\
  {\bf TOTAL} & {\sl 0.6} & -20.500 & -7.741 & -0.025 & -28.266 \\
  & {\sl 0.8} & -19.709 & -9.424 & -0.063 & -29.195 \\
  & {\sl 1.0} & -19.005 & -10.690 & -0.120 & -29.815 \\
\hline
\end{tabular}
\end{table}

\begin{table}
\caption{Values of $B_{AA}$ for $r=1$}
\begin{tabular}{ccrrrr}
  \hline  & Wilson & $O(a)$ impr. & $O(a^2)$ impr. & total \\
\hline  {\bf SAILS} & -2.353 & -3.568 & -5.582 & -11.502 \\
\hline  {\bf VERTEX} & 0.870 & -0.258 & 0.384 & 0.996 \\
\hline  {\bf $\half$ SELF-ENERGY} & -0.381 & -3.646 & 0 & -4.027 \\
\hline  {\bf operator TADPOLE} & & & & \\
  $+$ & -16.960 & 0 & 4.104 & -12.856 \\
  {\bf $\half$ leg TADPOLE} & & & & \\
\hline \hline {\bf TOTAL} & -18.824 & -7.472 & -1.093 & -27.389 \\
\hline
\end{tabular}
\end{table}

\begin{table}
\caption{Values of $B_{AB}$ for $r=1$}
\begin{tabular}{ccrrrr}
\hline  & Wilson & $O(a)$ impr. & $O(a^2)$ impr. & total \\
\hline  {\bf SAILS} & -0.953 & -1.183 & 2.391 & 0.255 \\
\hline  {\bf VERTEX} & 0.029 & -0.245 & 0.007 & -0.208 \\
\hline  {\bf $\half$ SELF-ENERGY} & 0 & 0 & 0 & 0 \\
\hline  {\bf operator TADPOLE} & & & & \\
  $+$ & 0 & 0 & -3.649 & -3.649 \\
  {\bf $\half$ leg TADPOLE} & & & & \\
\hline \hline {\bf TOTAL} & -0.924 & -1.428 & -1.251 & -3.603 \\
\hline
\end{tabular}
\end{table}

\begin{table}
\caption{Values of $B_{BA}$ for $r=1$}
\begin{tabular}{ccrrrr}
\hline  & Wilson & $O(a)$ impr. & $O(a^2)$ impr. & total \\
\hline  {\bf SAILS} & -3.108 & -4.731 & 5.465 & -2.373 \\
\hline  {\bf VERTEX} & 0.153 & -0.556 & -0.044 & -0.448 \\
\hline  {\bf $\half$ SELF-ENERGY} & 0 & 0 & 0 & 0 \\
\hline  {\bf operator TADPOLE} & & & & \\
  $+$ & 0 & 0 & -8.982 & -8.982 \\
  {\bf $\half$ leg TADPOLE} & & & & \\
\hline \hline {\bf TOTAL} & -2.955 & -5.287 & -3.562 & -11.803 \\
\hline
\end{tabular}
\end{table}

\begin{table}
\caption{Values of $B_{BB}$ for $r=1$}
\begin{tabular}{ccrrrr}
\hline  & Wilson & $O(a)$ impr. & $O(a^2)$ impr. & total \\
\hline  {\bf SAILS} & -6.421 & -3.899 & -4.313 & -14.633 \\
\hline  {\bf VERTEX} & 1.494 & -0.462 & 0.725 & 1.758 \\
\hline  {\bf $\half$ SELF-ENERGY} & -0.381 & -3.646 & 0 & -4.027 \\
\hline  {\bf operator TADPOLE} & & & & \\
  $+$ & -12.233 & 0 & 10.596 & -1.637 \\
  {\bf $\half$ leg TADPOLE} & & & & \\
\hline \hline {\bf TOTAL} & -17.540 & -8.006 & 7.009 & -18.538 \\
\hline
\end{tabular}
\end{table}

\begin{table}
\caption{Mixing matrix of the anomalous dimensions of the operators
$O_A$ and $O_B$}
\begin{tabular}{cc}
\hline
& ANOMALOUS DIMENSIONS \\
\hline
& \\
{\bf SAILS}
& $ \left[
\begin{array}{cc} 10/3 & 5/3 \\ 10/3 & 5
\\ \end{array} \right] $
\\
& \\
{\bf VERTEX}
& $ \left[
\begin{array}{cc} -1 & 1/3 \\ 2/3 & -2/3
\\ \end{array} \right] $
\\
& \\
$\half$ {\bf SELF-ENERGY}
& $ \left[
\begin{array}{cc} 2 & 0 \\ 0 & 2 \\ \end{array}
\right] $
\\
& \\
\hline \hline
& \\
{\bf TOTAL}
& $ \left[
\begin{array}{cc} 13/3 & 2 \\ 4 & 19/3 \\ \end{array}
\right] $
\\
& \\
\hline
\end{tabular}
\end{table}

\begin{table}
\caption{Mixing matrix for the $B$ constants
(the lattice values are for $r=1$)}
\begin{tabular}{cccc}
\hline
& Continuum $\overline{MS}$ & Lattice & Lattice \\
& & (Wilson) & (full improved) \\
\hline
& \\
{\bf SAILS}
& $ \left(
\begin{array}{cc} -\frac{31}{9} & -\frac{31}{18} \\ -\frac{31}{9}
& -\frac{93}{18} \\ \end{array} \right) $
& $ \left(
\begin{array}{cc} -2.353 & -0.953 \\ -3.108 & -6.421 \\ \end{array}
\right) $
& $ \left(
\begin{array}{cc} -11.502 & 0.255 \\ -2.373 & -14.633 \\ \end{array}
\right) $ \\
& \\
{\bf VERTEX}
& $ \left(
\begin{array}{cc} \frac{8}{9} & \frac{1}{9} \\ \frac{2}{9} & 1 \\
\end{array} \right) $
& $ \left(
\begin{array}{cc} 0.870 & 0.029 \\ 0.153 & 1.494 \\ \end{array}
\right) $
& $ \left(
\begin{array}{cc} 0.996 & -0.208 \\ -0.448 & 1.758 \\ \end{array}
\right) $ \\
& \\
$\half$ {\bf SELF-ENERGY}
& $ \left(
\begin{array}{cc} -1 & 0 \\ 0 & -1 \\ \end{array}
\right) $
& $ \left(
\begin{array}{cc} -0.381 & 0 \\ 0 & -0.381 \\ \end{array}
\right) $
& $ \left(
\begin{array}{cc} -4.027 & 0 \\ 0 & -4.027 \\ \end{array}
\right) $ \\
& \\
{\bf TADPOLES}
&
& $ \left(
\begin{array}{cc} -16.960 & 0 \\ 0 & -12.233 \\ \end{array}
\right) $
& $ \left(
\begin{array}{cc} -12.856 & -3.649 \\ -8.982 & -1.637 \\ \end{array}
\right) $ \\
& \\
\hline \hline
& \\
{\bf TOTAL}
& $ \left(
\begin{array}{cc} -\frac{32}{9} & -\frac{29}{18} \\ -\frac{29}{9}
& -\frac{93}{18} \\ \end{array} \right) $
& $ \left(
\begin{array}{cc} -18.824 & -0.924 \\ -2.955 & -17.540 \\ \end{array}
\right) $
& $ \left(
\begin{array}{cc} -27.389 & -3.603 \\ -11.803 & -18.538 \\ \end{array}
\right) $ \\
& \\
\hline
\end{tabular}
\end{table}

The results of our calculations are summarized in the Tables 5--12
presented in this section. The contributions coming from the standard
Wilson action (that is, the non-improved results), those coming from
the terms naively of order $a$ in the improvement procedure and those
coming from the terms naively of order $a^2$ are shown separately.
Contributions from different classes of diagrams, according
to the classification given in Appendix B, are also presented
separately. The accuracy of the numbers presented here depends only
on the accuracy of the numerical integration and it is better than 1\%.

In Table 5 we give the continuum $\overline{MS}$ renormalization
constants of the operators $O_{\{123\}}$, $O_A + O_B$ and
$O_{\{111\}}$. In Table 6 the results for the lattice renormalization
constants of the multiplicatively renormalizable operator
$O_{\{123\}}$ are reported. Our numbers for the Wilson
case agree, within errors, with those of Ref.~\cite{z3}.
In Tables 7--10 we give, for $r=1$ only, the values of the four
lattice $B$ constants relevant for the mixing between the operators
$O_A = O^q_{411} - \half (O^q_{422} + O^q_{433})$ and
$O_B = O^q_{141} + O^q_{114} - \half (O^q_{242} + O^q_{224} +
O^q_{343} + O^q_{334})$. In Table 11 for completeness we report
the mixing matrix of the anomalous dimensions of these
operators. In Table 12 a summary of the mixing matrices
for the $B$ constants both in continuum $\overline{MS}$ and on the
lattice (where we give separately the results for the Wilson case
and for the fully improved case) is given.

Two observations are in order here. First of all we notice
that for the physically relevant symmetric operator
$O^q_{\mathrm{DIS}} = \frac{1}{3} (O_A + O_B)$
the sum of the elements in each column of the lattice matrices
are not equal, signalling that mixing occurs. In the continuum
instead, as expected from the discussion of Sect.~2, it can be
easily checked from the numbers given in Tables 11 and 12 that
the combination $O_A + O_B$ is in fact multiplicatively renormalizable.
A second observation that emerges by looking at the results
of Table 12 is that, at least in this case, one should talk
of ``tadpole $+$ sails dominance'' rather than simply
of ``tadpole dominance'' in lattice perturbation theory.

We may summarize the results obtained in this paper
in the following formulae.

1) Standard Wilson action (Tables 6 and 12):

\begin{displaymath}
\widehat{O}^q_{\{123\}} =
\Bigg[ 1 - {\displaystyle\frac{25}{6 \pi^2 \beta}}
\log \mu a  - {\displaystyle\frac{1}{2 \pi^2 \beta}} B_{\{123\}}
\Bigg] O^q_{\{123\}}
\end{displaymath}
\begin{equation}
= \Bigg[ 1 - {\displaystyle\frac{0.422}{\beta}}
\log \mu a  + {\displaystyle\frac{0.963}{\beta}} \Bigg] O^q_{\{123\}},
\label{eq:rdivw}
\end{equation}

\begin{displaymath}
\widehat{O}^q_{\mathrm{DIS}} =
\frac{1}{3} \Bigg[ 1 - {\displaystyle\frac{25}{6 \pi^2 \beta}}
\log \mu a  - {\displaystyle\frac{1}{2 \pi^2 \beta}}
\Big( B_{AA} + B_{BA} \Big) \Bigg] O_A
\end{displaymath}
\begin{displaymath}
+ \frac{1}{3} \Bigg[ 1 - {\displaystyle\frac{25}{6 \pi^2 \beta}}
\log \mu a  - {\displaystyle\frac{1}{2 \pi^2 \beta}}
\Big( B_{AB} + B_{BB} \Big) \Bigg] O_B
\end{displaymath}
\begin{equation}
= \frac{1}{3} \Bigg[ 1 - {\displaystyle\frac{0.422}{\beta}}
\log \mu a  + {\displaystyle\frac{1.103}{\beta}} \Bigg] O_A
+ \frac{1}{3} \Bigg[ 1 - {\displaystyle\frac{0.422}{\beta}}
\log \mu a  + {\displaystyle\frac{0.935}{\beta}} \Bigg] O_B .
\label{eq:rugw}
\end{equation}

2) Nearest-neighbor improved action (Tables 6 and 12)

\begin{displaymath}
(\widehat{O}^q_{\{123\}})^{\mathrm{I}} =
\Bigg[ 1 - {\displaystyle\frac{25}{6 \pi^2 \beta}}
\log \mu a  - {\displaystyle\frac{1}{2 \pi^2 \beta}}
B_{\{123\}}^{\mathrm{I}} \Bigg] (O^q_{\{123\}})^{\mathrm{I}}
\end{displaymath}
\begin{equation}
= \Bigg[ 1 - {\displaystyle\frac{0.422}{\beta}}
\log \mu a  + {\displaystyle\frac{1.510}{\beta}} \Bigg]
(O^q_{\{123\}})^{\mathrm{I}} ,
\label{eq:rdivsw}
\end{equation}

\begin{displaymath}
(\widehat{O}^q_{\mathrm{DIS}})^{\mathrm{I}} =
\frac{1}{3} \Bigg[ 1 - {\displaystyle\frac{25}{6 \pi^2 \beta}}
\log \mu a  - {\displaystyle\frac{1}{2 \pi^2 \beta}}
\Big( B_{AA}^{\mathrm{I}} + B_{BA}^{\mathrm{I}} \Big) \Bigg]
(O_A)^{\mathrm{I}}
\end{displaymath}
\begin{displaymath}
+ \frac{1}{3} \Bigg[ 1 - {\displaystyle\frac{25}{6 \pi^2 \beta}}
\log \mu a  - {\displaystyle\frac{1}{2 \pi^2 \beta}}
\Big( B_{AB}^{\mathrm{I}} + B_{BB}^{\mathrm{I}} \Big) \Bigg]
(O_B)^{\mathrm{I}}
\end{displaymath}
\begin{displaymath}
= \frac{1}{3} \Bigg[ 1 - {\displaystyle\frac{0.422}{\beta}}
\log \mu a  + {\displaystyle\frac{1.985}{\beta}} \Bigg]
(O_A)^{\mathrm{I}}
\end{displaymath}
\begin{equation}
+ \frac{1}{3} \Bigg[ 1 - {\displaystyle\frac{0.422}{\beta}}
\log \mu a  + {\displaystyle\frac{1.122}{\beta}} \Bigg]
(O_B)^{\mathrm{I}} .
\label{eq:rugsw}
\end{equation}

Obviously physical quantities should at the end be independent from
the chosen regularization procedure. In particular the dependence from
the subtraction point, $\mu$, must disappear from physical hadronic
matrix elements. In fact, consistently to each order in perturbation
theory, the $\log \mu a$ terms get canceled in the product between the
Wilson coefficients and the matrix elements of the renormalized
operators that are eigenvectors of the anomalous dimension matrix.
The net result is that effectively the Wilson coefficients must be
taken at a momentum scale $a^{-1}$ and the operators turn out to be
effectively renormalized by ``reduced'' renormalization constants
$\widetilde{Z}$, obtained from the full expressions by dropping all
$\log \mu a$ terms.

Numerically for $N_c = 3$ at the typical values
$\beta \equiv 2 N_c / g_0^2 = 6$ and $r = 1$ and in the quenched
approximation, one gets for the unmixed quark operator
$O^q_{\{123\}}$ from Eqs.~(\ref{eq:rdivw}) and (\ref{eq:rdivsw})
\begin{center}
\begin{tabular}{rll}
$\widetilde{O}^q_{\{123\}} =$&$ 1.160~O^q_{\{123\}} $& Wilson case \\
$(\widetilde{O}^q_{\{123\}})^{\mathrm{I}}
=$&$ 1.252~(O^q_{\{123\}})^{\mathrm{I}}
$& Improved case ,
\end{tabular}
\end{center}
where by the superscript $~~\widetilde{}~~$ we mean that the
$\log \mu a$ term has been dropped from the expression
of the $Z$'s, as explained before.

Similarly in the case of the operator $O^q_{\mathrm{DIS}}$
one has from Eqs.~(\ref{eq:rugw}) and (\ref{eq:rugsw})
\begin{center}
\begin{tabular}{rll}
$\widetilde{O}^q_{\mathrm{DIS}} =$&$ \frac{1}{3} \left[
1.184~O_{A} + 1.156~O_{B} \right] $
& Wilson case \\
$(\widetilde{O}^q_{\mathrm{DIS}})^{\mathrm{I}}
=$&$ \frac{1}{3} \left[
1.331~(O_{A})^{I} + 1.187~(O_{B})^{\mathrm{I}} \right] $
& Improved case .
\end{tabular}
\end{center}

In the existing simulations corrections due to renormalization
and mixing effects were either calculated by means of tadpole
dominance \cite{Mart1,Mart2} or for simplicity neglected \cite{z1}.
We see that even this latter approximation is not quite adequate
because the effective renormalization constants and mixing
coefficients are not small. It should also be noted that
the renormalization constants do not seem to decrease
in magnitude with the order $N$ of the moment. This fact is already
evident in the Wilson case, where the renormalization constants
of the rank three operators are larger than those of the rank two
quark operators (see Ref.~\cite{first}), and it is confirmed by the
complete improved computation.

\appendix

\section{Perturbative expansion of the vertex operator
$(O^q_{\mu \nu \tau})^{\mathrm{I}}$}

In this appendix we present the perturbative expansion of the vertex
operator $(O^q_{\mu \nu \tau})^{\mathrm{I}}$,
given by Eq.~(\ref{eq:o3impr}). We have used the definitions
\begin{equation}
\stackrel{\displaystyle \rightarrow}{D}_{\mu} \psi_n = \frac{1}{2a}
\Big[ U_{n, \nu} \psi_{n + \mu} - U_{n - \mu, \mu}^{+} \psi_{n - \mu}
\Big]
\end{equation}
\begin{displaymath}
\overline{\psi} \stackrel{\displaystyle \leftarrow}{D}_{\mu}
= \frac{1}{2a} \Big[ \overline{\psi}_{n + \mu} U_{n, \nu}^{+}
 - \overline{\psi}_{n - \mu} U_{n - \mu, \mu} \Big] ,
\end{displaymath}
and the conventions
\begin{equation}
A_{n, \mu} = \int\! \frac{d^4 q}{(2 \pi)^4} \:
\e^{\displaystyle \mathrm{i} (q + q_{\mu} /2) n} A_{\mu}(q)
\end{equation}
\begin{equation}
\psi_n = \int\! \frac{d^4 q}{(2 \pi)^4} \:
\e^{\displaystyle \mathrm{i} q n} \psi (q)
\end{equation}
\begin{equation}
\overline{\psi}_n = \int\! \frac{d^4 q}{(2 \pi)^4} \:
\e^{\displaystyle - \mathrm{i} q n} \overline{\psi} (q) ,
\end{equation}
where the integrals are performed over the first Brillouin zone.
Throughout this appendix external and loop momenta are expressed
in lattice units.

Since the full Fourier transform of the operator
$(O^q_{\mu \nu \tau})^{\mathrm{I}}$ is very complicated,
we give here only the form it effectively takes when inserted
in the diagrams
of Fig.~B.2. Calling $p$ the external incoming and outgoing momentum,
and $k$ the fermion loop momentum, one finds, separating the various
contributions according to their naive order in $a$:

a) tree level

\begin{displaymath}
(O^q_{\mu \nu \tau})^{\mathrm{I}}(n = 0) |_{tree} =
- {\displaystyle\frac{1}{2}} \int_{-\pi}^{\pi} \frac{d^4 k}{(2 \pi)^4}
\overline{\psi}(k) \gamma_{\mu} {\displaystyle\frac{\sin k_{\nu}}{a}}
{\displaystyle\frac{\sin k_{\tau}}{a}}
\psi (k)
\end{displaymath}
\begin{equation}
+ {\displaystyle\frac{\mathrm{i} a r }{2}}
\int_{-\pi}^{\pi} \frac{d^4 k}{(2 \pi)^4} \overline{\psi}(k)
{\displaystyle\frac{\sin k_{\mu}}{a}}
{\displaystyle\frac{\sin k_{\nu}}{a}}
{\displaystyle\frac{\sin k_{\tau}}{a}}
\psi (k) \label{eq:q3tree}
\end{equation}
\begin{displaymath}
+ {\displaystyle\frac{a^2 r^2}{8}} \int_{-\pi}^{\pi}
\frac{d^4 k}{(2 \pi)^4}
\overline{\psi}(k) \sum_{\lambda,\lambda '}
\gamma_{\lambda} \gamma_{\mu} \gamma_{\lambda '}
{\displaystyle\frac{\sin k_{\lambda}}{a}}
{\displaystyle\frac{\sin k_{\lambda '}}{a}}
{\displaystyle\frac{\sin k_{\nu}}{a}}
{\displaystyle\frac{\sin k_{\tau}}{a}} \psi (k).
\end{displaymath}

b) order $g_0$

\begin{displaymath}
(O^q_{\mu \nu \tau})^{\mathrm{I}}(n = 0) |_{g_0} =
- {\displaystyle\frac{g_0}{2}} \int_{-\pi}^{\pi}
\frac{d^4 k}{(2 \pi)^4}
\overline{\psi}(p) \gamma_{\mu} \Big[
\cos ({\displaystyle\frac{k + p}{2}})_{\nu}
{\displaystyle\frac{\sin k_{\tau}}{a}} A_{\nu}(p-k)
\end{displaymath}
\begin{displaymath}
+ \cos ({\displaystyle\frac{k + p}{2}})_{\tau}
{\displaystyle\frac{\sin p_{\nu}}{a}} A_{\tau}(p-k)
\Big] \psi (k)
\end{displaymath}
\begin{displaymath}
+ {\displaystyle\frac{\mathrm{i} a g_0 r}{4}} \int_{-\pi}^{\pi}
\frac{d^4 k}{(2 \pi)^4} \overline{\psi}(p) \sum_{\lambda}
\Bigg\{ \gamma_{\mu} \gamma_{\lambda}
\Big[ \cos ({\displaystyle\frac{k + p}{2}})_{\nu}
{\displaystyle\frac{\sin k_{\tau}}{a}}
{\displaystyle\frac{\sin k_{\lambda}}{a}} A_{\nu}(p-k)
\end{displaymath}
\begin{displaymath}
+\cos ({\displaystyle\frac{k + p}{2}})_{\tau}
{\displaystyle\frac{\sin k_{\lambda}}{a}}
{\displaystyle\frac{\sin p_{\nu}}{a}} A_{\tau}(p-k)
\end{displaymath}
\begin{equation}
+\cos ({\displaystyle\frac{k + p}{2}})_{\lambda}
{\displaystyle\frac{\sin p_{\nu}}{a}}
{\displaystyle\frac{\sin p_{\tau}}{a}} A_{\lambda}(p-k) \Big]
\label{eq:q3g}
\end{equation}
\begin{displaymath}
+  \gamma_{\lambda} \gamma_{\mu} \Big[
\cos ({\displaystyle\frac{k + p}{2}})_{\nu}
{\displaystyle\frac{\sin k_{\tau}}{a}}
{\displaystyle\frac{\sin p_{\lambda}}{a}} A_{\nu}(p-k)
\end{displaymath}
\begin{displaymath}
+\cos ({\displaystyle\frac{k + p}{2}})_{\tau}
{\displaystyle\frac{\sin p_{\nu}}{a}}
{\displaystyle\frac{\sin p_{\lambda}}{a}} A_{\tau}(p-k)
\end{displaymath}
\begin{displaymath}
+\cos ({\displaystyle\frac{k + p}{2}})_{\lambda}
{\displaystyle\frac{\sin k_{\nu}}{a}}
{\displaystyle\frac{\sin k_{\tau}}{a}} A_{\lambda}(p-k) \Big]
\Bigg\} \psi (k)
\end{displaymath}
\begin{displaymath}
+ {\displaystyle\frac{a^2 g_0 r^2}{8}} \int_{-\pi}^{\pi}
\frac{d^4 k}{(2 \pi)^4}
\overline{\psi}(p) \sum_{\lambda,\lambda '}
\gamma_{\lambda} \gamma_{\mu} \gamma_{\lambda '}
\Bigg[ \cos ({\displaystyle\frac{k + p}{2}})_{\nu}
{\displaystyle\frac{\sin p_{\lambda '}}{a}}
{\displaystyle\frac{\sin k_{\tau}}{a}}
{\displaystyle\frac{\sin k_{\lambda}}{a}} A_{\nu}(p-k)
\end{displaymath}
\begin{displaymath}
+\cos ({\displaystyle\frac{k + p}{2}})_{\tau}
{\displaystyle\frac{\sin p_{\nu}}{a}}
{\displaystyle\frac{\sin p_{\lambda '}}{a}}
{\displaystyle\frac{\sin k_{\lambda}}{a}} A_{\tau}(p-k)
\end{displaymath}
\begin{displaymath}
+\cos ({\displaystyle\frac{k + p}{2}})_{\lambda}
{\displaystyle\frac{\sin k_{\nu}}{a}}
{\displaystyle\frac{\sin k_{\tau}}{a}}
{\displaystyle\frac{\sin k_{\lambda '}}{a}} A_{\lambda}(p-k)
\end{displaymath}
\begin{displaymath}
+ \cos ({\displaystyle\frac{k + p}{2}})_{\lambda '}
{\displaystyle\frac{\sin p_{\nu}}{a}}
{\displaystyle\frac{\sin p_{\tau}}{a}}
{\displaystyle\frac{\sin p_{\lambda '}}{a}} A_{\lambda '}(p-k)
\Bigg] \psi (k).
\end{displaymath}

The above formula is given in the kinematical configuration in which
the incoming gluon momentum lands on the incoming quark leg.
If the gluon is attached to the outgoing quark leg, one must exchange
$p$ and $k$. Notice that the first lines in Eqs.~(\ref{eq:q3tree})
and (\ref{eq:q3g}) correspond to the non-improved expression
of the operator.

We do not give here the expression of the $O(g_0^2)$ terms because,
besides being extremely complicated, they are not actually necessary
for our computation. In fact we need them only when either the gluon
or the quark legs are contracted to make a tadpole loop, and in this
situation the tadpole directly factorizes out.

The case in which two or more indices are identical is easily
derived from the expressions above.

\section{Diagrams}

\begin{figure}
\begin{picture}(40000,15000)
\drawline\fermion[\NE\REG](10000,2000)[3250]
\drawline\gluon[\E\REG](\pbackx,\pbacky)[9]
\drawline\fermion[\NE\REG](\fermionbackx,\fermionbacky)[6750]
\drawline\fermion[\NW\REG](\gluonbackx,\gluonbacky)[6750]
\put(\fermionbackx,\fermionbacky){\circle*{800}}
\drawline\fermion[\SE\REG](\gluonbackx,\gluonbacky)[3250]
\end{picture}
\caption{The graph that symbolically represents the 1-loop vertex
correction to the insertion of the $(O^q_{\mu \nu})^{\mathrm{I}}$
operator. The operator insertion is indicated by a dot.
The wavy line is a gluon.}
\end{figure}
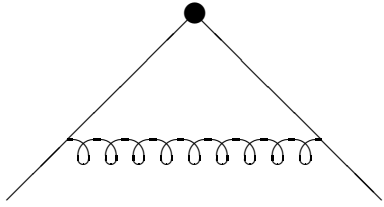

\begin{figure}
\begin{picture}(40000,15000)
\put(0,0){(a) Vertex}
\drawline\fermion[\NE\REG](0,2000)[3250]
\drawline\gluon[\E\REG](\pbackx,\pbacky)[9]
\drawline\fermion[\NE\REG](\fermionbackx,\fermionbacky)[6750]
\put(\fermionbackx,\fermionbacky){\circle*{800}}
\drawline\fermion[\NW\REG](\gluonbackx,\gluonbacky)[6750]
\drawline\fermion[\SE\REG](\gluonbackx,\gluonbacky)[3250]
\put(20000,0){(b) Sail}
\drawline\fermion[\NE\REG](20000,2000)[6000]
\drawloop\gluon[\NW 5](\pbackx,\pbacky)
\drawline\fermion[\NE\REG](\fermionbackx,\fermionbacky)[4000]
\put(\fermionbackx,\fermionbacky){\circle*{800}}
\drawline\fermion[\SE\REG](\fermionbackx,\fermionbacky)[10000]
\end{picture}
\begin{picture}(40000,15000)
\put(0,0){(c) Sail}
\drawline\fermion[\NE\REG](0,2000)[10000]
\put(\fermionbackx,\fermionbacky){\circle*{800}}
\drawloop\gluon[\NE 5](\pbackx,\pbacky)
\drawline\fermion[\SE\REG](\fermionbackx,\fermionbacky)[10000]
\put(20000,0){(d) Operator tadpole}
\drawline\fermion[\NE\REG](20000,2000)[10000]
\put(\fermionbackx,\fermionbacky){\circle*{800}}
\global\advance\pbackx by -400
\drawloop\gluon[\NW 8](\pbackx,\pbacky)
\drawline\fermion[\SE\REG](\fermionbackx,\fermionbacky)[10000]
\end{picture}
\begin{picture}(40000,15000)
\put(0,0){(e) Self-energy}
\drawline\fermion[\NE\REG](0,2000)[3000]
\drawloop\gluon[\NW 5](\pbackx,\pbacky)
\drawline\fermion[\NE\REG](\fermionbackx,\fermionbacky)[7000]
\put(\fermionbackx,\fermionbacky){\circle*{800}}
\drawline\fermion[\SE\REG](\fermionbackx,\fermionbacky)[10000]
\put(20000,0){(f) Self-energy}
\drawline\fermion[\NE\REG](20000,2000)[10000]
\put(\fermionbackx,\fermionbacky){\circle*{800}}
\drawline\fermion[\SE\REG](\fermionbackx,\fermionbacky)[3000]
\drawloop\gluon[\NE 5](\pbackx,\pbacky)
\drawline\fermion[\SE\REG](\fermionbackx,\fermionbacky)[7000]
\end{picture}
\begin{picture}(40000,15000)
\put(0,0){(g) Leg tadpole}
\drawline\fermion[\NE\REG](0,2000)[5000]
\global\advance\pbacky by 500
\drawloop\gluon[\SW 8](\pbackx,\pbacky)
\drawline\fermion[\NE\REG](\fermionbackx,\fermionbacky)[5000]
\put(\fermionbackx,\fermionbacky){\circle*{800}}
\drawline\fermion[\SE\REG](\fermionbackx,\fermionbacky)[10000]
\put(20000,0){(h) Leg tadpole}
\drawline\fermion[\NW\REG](35000,2000)[5000]
\global\advance\pbacky by -1500
\global\advance\pbackx by 2000
\drawloop\gluon[\NW 8](\pbackx,\pbacky)
\drawline\fermion[\NW\REG](\fermionbackx,\fermionbacky)[5000]
\put(\fermionbackx,\fermionbacky){\circle*{800}}
\drawline\fermion[\SW\REG](\fermionbackx,\fermionbacky)[10000]
\end{picture}
\caption{The different types of graphs contributing to the
1-loop approximation of the matrix element
$\langle p| (O^q_{\mu \nu \tau})^{\mathrm{I}} |p \rangle$.}
\end{figure}

In what follows we show the 1-loop diagrams that we have
calculated for each one of the operators discussed in
this work. They are accompanied by a label that allows an
easy connection with the Tables presented in Sect.~5.

Actually, each graph in this appendix corresponds to several diagrams
in the perturbative expansion. For example, the vertex correction
of Fig.~B.1 symbolically represents the sum of 12 different diagrams.
They come out remembering that for the quark-gluon interaction one can
either take the standard Wilson vertex of Eq.~(\ref{eq:standWil})
or the improved vertex of Eq.~(\ref{eq:newimpr}), and for the operator
insertion either its unrotated form or the $O(a)$ or else the $O(a^2)$
corrections (see Eqs.~(\ref{eq:o3impr}) and (\ref{eq:q3tree})).

We report in Fig.~B.2 the types of graphs that contribute
to the 1-loop calculation of the matrix element
$\langle p| (O^q_{\mu \nu \tau})^{\mathrm{I}} |p \rangle$,
with the understanding that the drawings we show are representative
of the structure of full sets of improved diagrams.

\section{$\alpha$-integration}

To reduce the computing time in the numerical evaluation
of the integrals that involve fermion propagators, we have chosen to
perform analytically the integration over the Feynman parameter
$\alpha$, and to leave for the numerical integration only a
four-dimensional expression. To this end after the subtraction
of the standard logarithmic divergence we need to compute
integrals of the form
\begin{equation}
{\mbox{\large F}}_{n m} \bigg( f(k),g(k)\bigg) = \int_{0}^{1} \!
d\alpha \: \frac{\alpha^{n}}{[f(k) + \alpha g(k)]^{m}} .
\end{equation}
In our calculations we explicitly have
\begin{eqnarray}
g(k) & = & 4 \sum_{\lambda} \sin^2 \frac{k_{\lambda}}{2}
- \Big[ \sum_{\lambda} \sin^2 k_{\lambda} + 4 r^2 (\sum_{\lambda}
\sin^2 \frac{k_{\lambda}}{2})^2 \Big] \nonumber \\
f(k) & = & \Big[ \sum_{\lambda} \sin^2 k_{\lambda}
+ 4 r^2 (\sum_{\lambda} \sin^2 \frac{k_{\lambda}}{2})^2 \Big] .
\end{eqnarray}
The functions ${\mbox{\large F}}_{n m}$ satisfy the recurrence
relations
\begin{equation}
\frac{\partial}{\partial f} {\mbox{\large F}}_{n m} =
- m \cdot {\mbox{\large F}}_{n\ m+1}
\end{equation}
\begin{equation}
\frac{\partial}{\partial g} {\mbox{\large F}}_{n m} =
- m \cdot {\mbox{\large F}}_{n+1\ m+1}
\end{equation}
which make simpler their computation and allow for a check of the
formulae given below.

The formulae needed in this work are:

\begin{eqnarray}
{\mbox{\large F}}_{0 2} & = & \frac{1}{f \cdot (g+f)} \nonumber \\
{\mbox{\large F}}_{1 2} & = & \frac{1}{g} \cdot \left[ \frac{1}{g}
\log \left( 1+\frac{g}{f} \right) -\frac{1}{g+f} \right] \nonumber \\
{\mbox{\large F}}_{2 2} & = & \frac{1}{g^2} \cdot \left[ 1-2 \frac{f}{g}
\log \left( 1+\frac{g}{f} \right) +\frac{f}{g+f} \right] \nonumber \\
{\mbox{\large F}}_{3 2} & = & \frac{1}{g^2}\cdot \left[ \frac{1}{2}
-2 \frac{f}{g} +3 \frac{f^2}{g^2} \log \left( 1+\frac{g}{f} \right) -
\frac{f^2}{g\cdot (g+f)} \right] \nonumber \\
{\mbox{\large F}}_{4 2} & = & \frac{1}{g^2}\cdot \left[ \frac{1}{3}
-\frac{f}{g} +3 \frac{f^2}{g^2} -4 \frac{f^3}{g^3}
\log \left( 1+\frac{g}{f} \right)
+ \frac{f^3}{g^2\cdot (g+f)} \right] \nonumber \\
{\mbox{\large F}}_{0 3} & = & \frac{g+2 f}{2 f^2 \cdot (g+f)^2}
\nonumber \\
{\mbox{\large F}}_{1 3} & = & \frac{1}{2 f\cdot (g+f)^2} \nonumber \\
{\mbox{\large F}}_{2 3} & = & \frac{1}{g^2}\cdot \left[ \frac{1}{g}
\log \left( 1+\frac{g}{f} \right) -\frac{3 g+2 f}{2\cdot (g+f)^2}
\right] \nonumber \\
{\mbox{\large F}}_{3 3} & = & \frac{1}{g^3} \cdot \left[ 1-3 \frac{f}{g}
\log \left( 1+\frac{g}{f} \right) +f \frac{5 g+4 f}{2\cdot (g+f)^2}
\right] \nonumber \\
{\mbox{\large F}}_{4 3} & = & \frac{1}{g^3} \cdot \left[ \frac{1}{2}
- 3 \frac{f}{g} + 6 \frac{f^2}{g^2} \log \left( 1+\frac{g}{f} \right)
-f^2 \frac{7 g+6 f}{2 g \cdot (g+f)^2} \right] \nonumber \\
{\mbox{\large F}}_{5 3} & = & \frac{1}{g^3} \cdot \left[ \frac{1}{3}
- \frac{3}{2} \frac{f}{g} + 6 \frac{f^2}{g^2}
- 10 \frac{f^3}{g^3} \log \left( 1+\frac{g}{f} \right)
+f^3 \frac{9 g+8 f}{2 g^2 \cdot (g+f)^2} \right] \nonumber \\
{\mbox{\large F}}_{6 3} & = & \frac{1}{g^3} \cdot \left[ \frac{1}{4}
- \frac{f}{g} + 3 \frac{f^2}{g^2} - 10 \frac{f^3}{g^3}
+ 15 \frac{f^4}{g^4} \log \left( 1+\frac{g}{f} \right)
-f^4 \frac{11 g+10 f}{2 g^3 \cdot (g+f)^2} \right] \nonumber \\
{\mbox{\large F}}_{0 4} & = & \frac{g^2+3 g f+3 f^2}{3 f^3
\cdot (g+f)^3} \nonumber \\
{\mbox{\large F}}_{1 4} & = & \frac{g+3 f}{6 f^2 \cdot (g+f)^3} \\
{\mbox{\large F}}_{2 4} & = & \frac{1}{3 f\cdot (g+f)^3} \nonumber \\
{\mbox{\large F}}_{3 4} & = & \frac{1}{g^3} \cdot \left[ \frac{1}{g}
\log \left( 1+\frac{g}{f} \right) -\frac{11 g^2+15 g f+6 f^2}{6\cdot
(g+f)^3} \right] \nonumber \\
{\mbox{\large F}}_{4 4} & = & \frac{1}{g^4} \cdot \left[ 1-4 \frac{f}{g}
\log \left( 1+\frac{g}{f} \right) +f \frac{13 g^2+21 g f +9 f^2}
{3\cdot (g+f)^3} \right] \nonumber \\
{\mbox{\large F}}_{2 5} & = & \frac{g + 4 f}{12 f^2 \cdot (g + f)^4}
\nonumber \\
{\mbox{\large F}}_{3 5} & = & \frac{1}{4 f \cdot (g + f)^4} \nonumber \\
{\mbox{\large F}}_{4 5} & = & \frac{1}{g^4} \cdot \left[ \frac{1}{g}
\log \left( 1+\frac{g}{f} \right) - \frac{25 g^3 + 52 g^2 f + 42 g f^2
+ 12 f^3} {12 \cdot (g + f)^4} \right] \nonumber \\
{\mbox{\large F}}_{5 5} & = & \frac{1}{g^5} \cdot \left[ 1 - 5
\frac{f}{g} \log \left( 1 + \frac{g}{f} \right) + f \frac{77 g^3
+ 188 g^2 f + 162 g f^2 + 48 f^3}{12 \cdot (g + f)^4} \right]
\nonumber .
\end{eqnarray}

\begin{ack}
We wish to thank M.G\"ockeler, H.Perlt and G.Schierholz for
communicating to us their results prior to publication, and
for many useful discussions.
We also thank M. Ciuchini, E. Franco and M. Testa for comments
and discussions.
\end{ack}

\end{document}